# Validation of a model for estimating the strength of the vortex created by a Vortex Generator from its Bound Circulation

**Martin O.L.Hansen [1*], Antonios Charalampous[1], Jean-Marc Foucaut[2,3], Christophe Cuvier[2,3], Clara M. Velte[4]**

[1*] Department of Wind Energy, Technical University of Denmark, 2800, Kgs. Lyngby, Denmark ; molh@dtu.dk

[2] Ecole Centrale Lille,Villeneuve d'Asq, France ; christophe.cuvier@centralelille.fr

[3] Université Lille Nord de France, LMFL-Laboratoire de Mécanique des Fluides de Lille, France ; jean-marc.foucaut@ec-lille.fr

[4] Department of Mechanical Engineering, Technical University of Denmark, 2800, Kgs. Lyngby; cmve@dtu.dk

**Abstract:** A hypothesis is tested and validated for predicting the vortex strength induced by a vortex generator in wall-bounded flow by combining the knowledge of the Vortex Generator (VG) geometry and the approaching boundary layer velocity distribution. In this paper, the spanwise distribution of bound circulation on the vortex generator is computed from integrating the pressure force along the VG height calculated using CFD. It is then assumed that all this bound circulation is shed into the wake to fulfill Helmholtz's theorem and then curl up into one primary tip vortex. To validate this, the trailed circulation estimated from the distribution of the bound circulation is compared to the one in the wake behind the vortex generator determined directly from the wake velocities at some downstream distance. In practical situations, the pressure distribution on the vane is unknown and consequently other estimates of the spanwise force distribution on the VG must instead be applied, such as using 2D airfoil data corresponding to the VG geometry at each wall-normal distance. Such models have previously been proposed and used as an engineering tool to aid preliminary VG design and it is not the purpose of this paper to refine such engineering models, but to validate their assumptions such as applying a lifting line model on a VG that has a very low aspect ratio and placed in wall boundary layer. Herein, high Reynolds number boundary layer measurements of VG induced flow were used to validate the Reynolds Averaged Navier-Stokes (RANS) modeled circulation results and are used for further illustration and validation of the hypothesis.

**Keywords:** Vortex Generators, Turbulent boundary layer flow control, bound circulation, trailed circulation

**Abbreviations:**

| | |
|---|---|
| VG | Vortex Generator |
| RANS | Reynolds Averaged Navier Stokes |
| LOV | Lamb-Oseen Vortex |
| PIV | Particle Image Velocimetry |
| RST | Reynolds-Stress Transport Turbulence Model |
| S-A | Spalart-Allmaras Transport Turbulence Model |

## 1. Introduction

Vortex generators (VGs), as first described by Taylor in the late 1940s [1-3] and originally used to suppress flow separation in diffusors, are today commonly used aerodynamic devices for wind turbine blades, mainly to delay separation at high angles of attack. This increases the maximum lift coefficient and thus reduces the necessary chord for the same lift, resulting in more slender and cost-effective blades [4]. VGs are also common in e.g. heat exchangers [5], solar receivers [6], thermoelectric power generators [7] and even in microchannels with nanofluids [8], to name but a few examples.

VGs in principle consist of small vanes, typically with a triangular or rectangular geometry. They are mounted in the boundary layer with an angle of incidence to the incoming flow that creates vortices behind the VGs causing momentum exchange between the region near the wall and the outer parts of the boundary layer [1-3]. Therefore, the boundary layer becomes more resistant to flow reversal in adverse pressure gradients.

The flow behind VGs was investigated experimentally using Particle Image Velocimetry (PIV) in Velte et al. [9] and it was shown that the generated primary vortices possess helical symmetry within a specific range of device angles. Many engineering-based models have been proposed over the years for predicting the primary (tip) vortex circulation, see e.g. [10-15]. The models are typically of a design where the circulation would be proportional to the product between the vortex generator height and the average streamwise velocity at the upper tip of the vortex generator. However, these kinds of models typically do not take the variable vortex generator geometry and boundary layer profile into consideration. Furthermore, it has been observed that the occurrence of secondary vortices (from spanwise boundary layer separation and the roll-up effect around the vane base) can substantially affect the circulation of the primary vortex [16]. A more correct manner to obtain the true primary vortex strength (circulation) is thus hypothesized to be possible through the integration of the bound vorticity, which was previously tested in [17].

Further, the generated flow can be investigated with the use of Computational Fluid Mechanics (CFD) modeling. A widely used method is the so-called BAY model [18] where body forces acting on the air near the VGs are specified to induce a realistic flow-field. However, a numerical model that directly resolves the VG geometry should be more accurate since no prior assumptions on the body forces are required, as has been done, e.g. in [19], and the results are to some success compared to both experiments and the BAY model.

In the current work, CFD simulations are performed on a spanwise row of rectangular and triangular VGs. The objective is to test the hypothesis that the strength of the produced vortex responsible for the mixing, quantified as the circulation $\Gamma$, can be calculated from the distribution of the bound circulation along the span of the VG. The local bound circulation is determined from the local sectional lift. The circulation of the streamwise vortex can also be directly computed from the flowfield behind the VG with the direct use of its definition and with the use of the Lamb-Oseen Vortex (LOV). This is then compared to the one estimated from the bound circulation distribution and if this hypothesis is valid, then it will be possible to use this information in preliminary VG designs.

## 2. Methods

### 2.1 Experimental setup

The measurements were carried out in the high Reynolds number wind tunnel in the Laboratoire de Mécanique des Fluides de Lille (LMFL) wind tunnel facility, see Figure 1. This wind tunnel was designed to generate a fully developed turbulent boundary layer with a thickness of $\delta = 30$ cm inside the 20 m long test section. The resulting Reynolds number based on the momentum thickness was 17000. The large dimensions of the boundary layer allowed for high effective resolution of the studied flow scales as well as matching of the Reynolds number to that of VG generated flows on actual wind turbine blades, $Re_h = U_\infty h / \nu = 8.5 \text{ ms}^{-1} \cdot 0.06 \text{ m} / 15 \cdot 10^{-6} \text{ m}^2\text{s}^{-1} = 34\,000$, where $U_\infty = 8.5 \text{ ms}^{-1}$ is the freestream velocity and $h = 0.06$ m is the VG height.

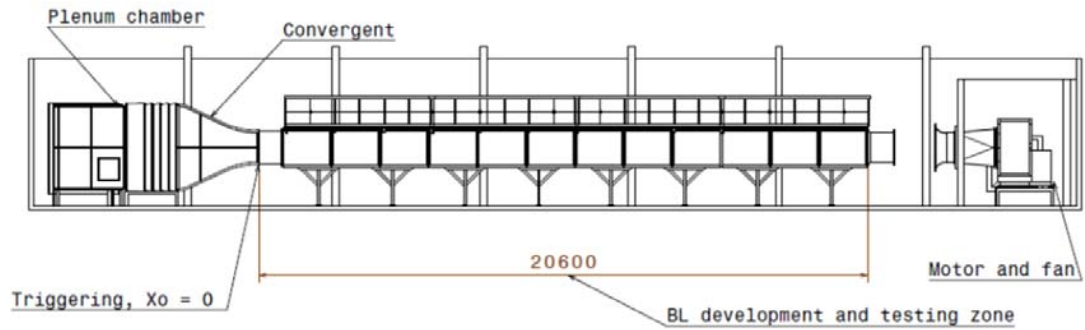

Figure 1 A sketch of the LMFL wind tunnel (figure originally extracted from http://lml.univ-lille1.fr/?page=144&menu_curr_set=144).

Passive VGs of triangular and rectangular shape, respectively, were introduced into the boundary layer at a stage where the boundary layer of 30 cm thickness only developed at a rate of 1 mm per meter in the downstream direction. Both VG geometries were of height h=60 mm and length l=2h=120 mm and mounted with an angle of incidence of 18° to the incoming mean flow direction. A top view of the configuration is illustrated in Figure 2, where L=2.5h is the distance between the trailing edges in a VG pair and λ=6h is the spanwise period. This geometric setting, producing counter-rotating vortices, is identical for the triangular and rectangular VG experiment configurations.

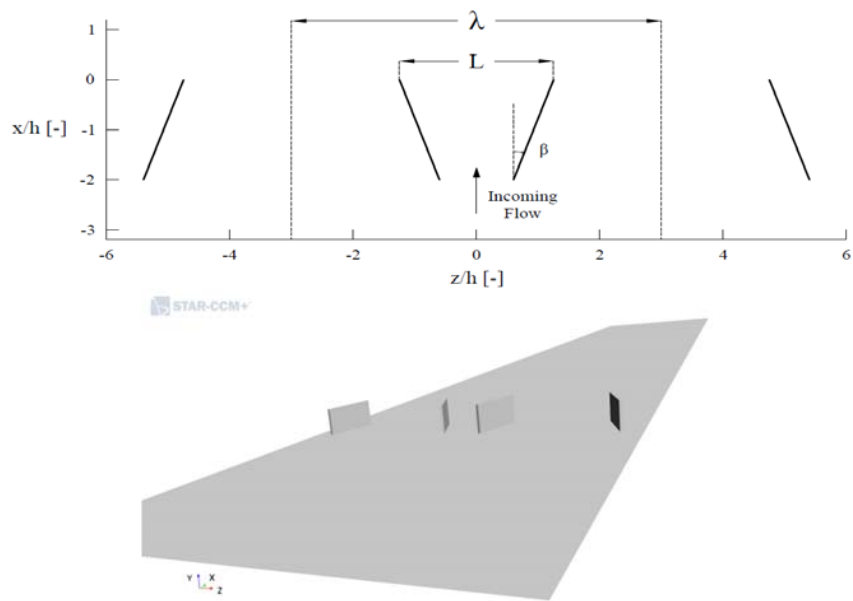

Figure 2 Top view of the configuration design showing the geometric parameters (top). A 3-D drawing of the set-up (bottom).

The flow was imaged in a cross-plane using a laser sheet illuminating smoke particles that could faithfully trace the flow, see Figure 3. The laser used is a BMI YAG laser with dual cavities, producing energies of 220 mJ per pulse at 12 Hz repetition frequency. The light sheet was introduced from above. A mirror below the transparent bottom wall, where the VGs were mounted, was used to utilize further the light intensity from the widely spread laser sheet. The beam waist was located approximately at the wall, with a width of approximately 1 mm (measured from burning paper). Furthermore, the light sheets of the two laser pulses were separated by approximately 1.5 mm in the downstream direction, to increase the dynamic range of the measured velocities. The seeding consisted of polyethylene glycol particles with a diameter of the order of 1 μm. This plane was located at variable distances x/h downstream of the VGs. The setup also consisted of four Hamamatsu 2k x 2k pixel cameras equipped with Nikkor 105 mm lenses at f# 5.6. The cameras were at a 45° angle to the light sheet and were therefore equipped also with adjustable Scheimpflug mounts and adjusted accordingly.

The cameras were arranged as two separate stereoscopic PIV setups, where each field of view covered one pair of vortex generators and the fields were ensured to be slightly overlapping to be able to merge the fields into a common wide field of view of about 72 x 28 cm² (corresponding to approximately 2.3δ x δ) [20,21]. The cameras were placed outside of the wind-tunnel.

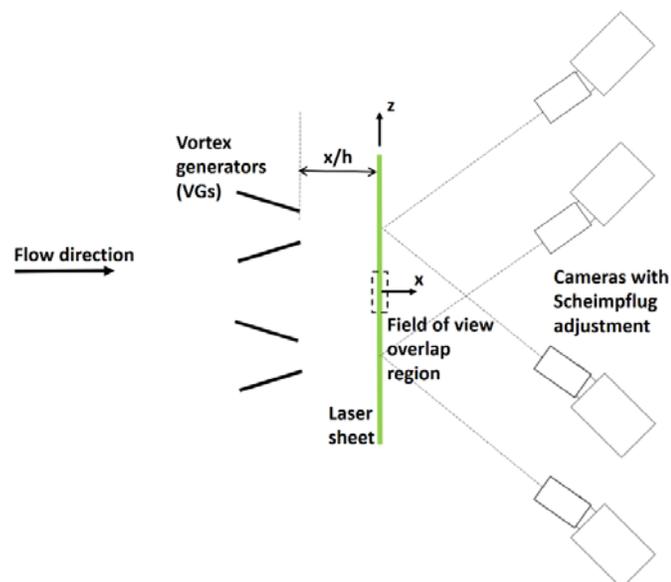

Figure 3 Schematic of the Stereoscopic Image Velocimetry setup.

To extract velocity fields from the images, the in-house PIV processing code was employed. The image processing was conducted over an interrogation area size of 24x32 pixels with an overlap of 60%, where the non-uniform aspect ratio stems from the stretching of the field from the camera perspective angle. In the processing, the self-correction of the calibration due to misalignment between the calibration target and the light sheet has been properly accounted and self-corrected for. The spacing between the interrogation areas after processing was 2 mm, corresponding to 40 wall units. The accuracy of the resulting velocity measurements has been thoroughly analyzed in [21], yielding approximately 3% for y/δ larger than approximately 0.1 and not exceeding approximately 12% in the near-wall region across all three velocity components.

An example of the ensemble averaged three-component velocity field measured downstream of the triangular vanes can be seen in Figure 4.

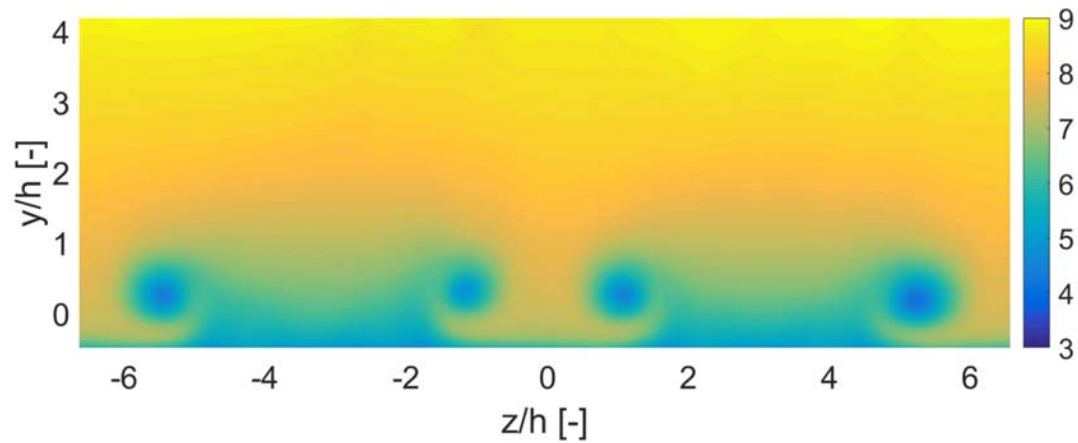

Figure 4 Example of the ensemble averaged velocity field at 3h downstream of the rectangular VGs.

*2.2 Numerical model*

Based on the experimental setup presented in [20, 21], two RANS models and a Reynold stress model that fully resolve the rectangular and triangular VGs geometries are developed using the commercial software STAR – CCM+ [22]. The geometry parts are created and assembled within the software's environment with the use of the integrated CAD tool. The authors' previous experience on modeling vortex generators with models of varying degrees of complexity motivated the inclusion of the Reynolds stress model [23-24].

The tunnel free-stream velocity U is set as the inlet condition to 8 m/s. The distance between the velocity inlet boundary condition and VGs' leading edges is set to 3.72 m. A parametric analysis using various lengths is conducted for the selection of this value ensuring a satisfying development of the velocity profile noting that the VGs are placed deeply in the boundary layer. The distance between the trailing edges and the rear downstream plane where a pressure outlet boundary condition is applied, is set equal to 3.2 m. This distance is chosen so that the flow does not reverse on the pressure outlet as the flow in the wake is undisturbed. The distance between the symmetry plane boundary conditions and the trailing edges is 0.075 m, i.e. half the distance between the trailing edges of the middle VG pair (see Figure 2). The height of the computational domain is set equal to 0.4 m. The VGs as well as the horizontal plane where the vanes are mounted are treated as wall boundaries. The Turbulence Intensity is specified as 1% and the Turbulence Viscosity Ratio is equal to 10. The mesh discretization is 11.3 $\cdot 10^6$ and 8.2 $\cdot 10^6$ polyhedral cells for the rectangular and triangular set-ups respectively. The polyhedral meshing option is chosen since it provides a detailed and balanced solution for complicated studies. Further, it requires less computational power since it generates five times fewer cells than the equivalent tetrahedral. Another advantage is that it does not require more surface preparation [22]. The characteristic dimension called Base size was set equal to 0.05 m for generating the mesh discretization of the overall domain. Specifically, for the rectangular configuration, two relative sizes were chosen for refining the areas that required detailed resolution. A relative size is chosen for discretizing the region close to the VGs and a second refined region is modelled for capturing the far wake region. A different approach was implemented for discretizing the triangular geometry. The area close to the VGs was only refined since the experimental data used for triangular model's validation exist only up to 12h downstream planes of the VGs. For the rectangular configuration, the validation sections were up to 50h downstream (i.e. further into the wake). Twenty prism layers are attached near the boundaries for enhanced capturing of the boundary layer development. An illustration of these layers implemented in the rectangular VGs can be seen in

Figure 5 (top). The final mesh discretization of the rectangular VGs model can be in

Figure 5 (bottom). An overview of the parameters used to generate the two grids is presented in **Error! Reference source not found.** below.

| Polyhedral mesh | Rectangular | Triangular |
|---|---|---|
| Base size [m] | 0.05 | 0.05 |
| Relative size close to VGs [%] | 9 | |
| Relative size in the wake region [%] | 11 | - |
| Prism layer mesh | ReD | TrD |
| Number of prism layers | 20 | 20 |
| Prism layer stretching | 1.5 | 1.5 |
| Prism layer thickness [% of the base size] | 5 | 5 |
| Total amount of cells | $11.3 \cdot 10^6$ | $8.2 \cdot 10^6$ |

**Table 1 Parameters used for the mesh discretization of the rectangular and triangular VGs.**

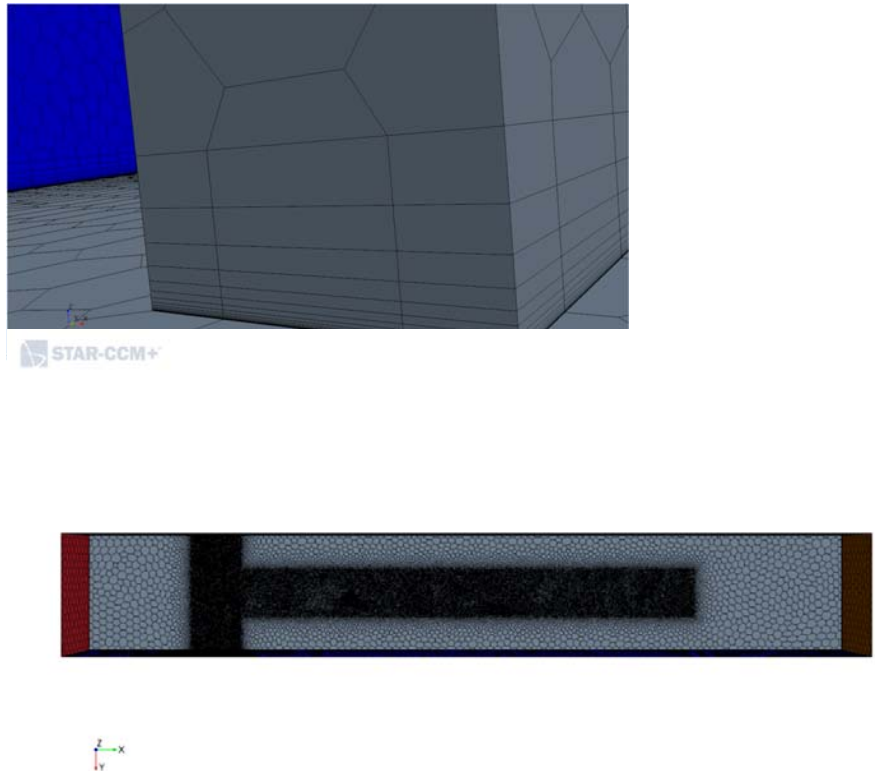

Figure 5 Prism layers attached to the rectangular VGs (top). Top view of the mesh discretization of the rectangular VGs (bottom).

Due to the symmetry of the configuration, the downstream region behind the middle pair is the only one refined in detail as it is seen in

Figure 5. This practice reduced significantly the computational time and allowed to achieve the highest discretization in the confined area. A sensitivity analysis was carried out for achieving the optimum quality of the mesh discretization given the computational resources. It was observed that the solutions remained almost constant when lower, compared to the aforementioned, relative sizes were applied (i.e. higher resolution), while the computational time was significantly increased. The three Transport Turbulence Models used to close the RANS equations were Reynolds-Stress, k-$\omega$ SST and Spalart-Allmaras. The solution converge criteria were set to 4000 iteration steps with a tolerance limit equal to 1e-6. The convergence of the solution was also verified by monitoring the behavior of the residuals. For the rectangular configuration, the residuals reached the tolerance limit and remained almost constant after the first 2000 iteration steps, while for the triangular set up 1200 iteration steps needed to meet the same limit. To validate the accuracy of the numerical models, the results are compared quantitatively with the experimental results using the circulation of the velocity fields as an indicator. The y+ values on the VG surface using the Reynolds Stress model which is the one used for the final comparison of the results varied between 3 and 10, lowest at the leading edge and increasing towards the trailing edge. The circulation is estimated with the direct use of path integration, Eq. (8) and the Lamb - Oseen Vortex, Eq. (7).

## 3. Estimating the bound and trailed circulation from CFD

Even though the VG aspect ratio is low, it is still assumed that the flow is locally 2-D with only a small spanwise component and thereby the bound circulation at a given height can be estimated using the Kutta-Joukowski theorem from 2-D airfoil theory:

$$l(y) = \rho u_x(y)\Gamma_b(y) \tag{1}$$

where y is the normal distance from the root of the VG, $\rho$ is the fluid density and $\Gamma_b(y)$ is the bound circulation at each location. The bound circulation $\Gamma_b(y)$ thus varies with the VG geometry and the incoming flow as a function of the wall normal distance y. The velocity profile shown in Figure 6 corresponds to the computed streamwise average velocity profile at 3h upstream i.e. x/h=-3h, where the flow is undisturbed by the vanes.

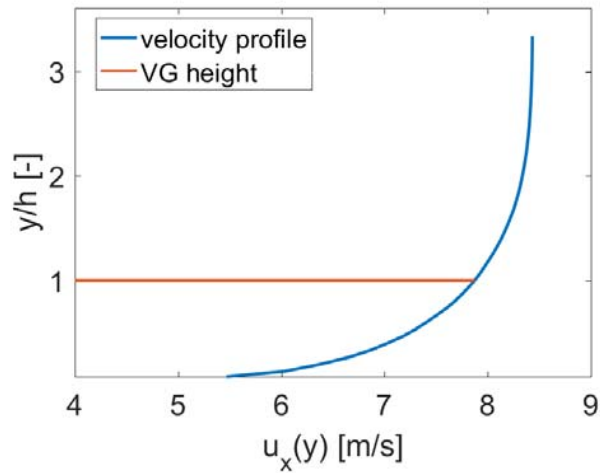

Figure 6 Streamwise average velocity profile at 3h upstream, RST turbulence model.

First, the pressure load normal to the VG surface [N/m] is found from integrating the pressure jump at the height y from the leading edge, s=0, to trailing edge, s=c(y), of the VG as shown in Figure 7.

$$f_n(y) = \int_{s=0}^{s=c(y)} \Delta p(s,y) ds \tag{2}$$

Since the pressure force acts perpendicularly to the VG surface, the lift force is estimated by projecting this with the geometric pitch angle, $\beta$, of the VG to the incoming flow as

$$l(y) = f_n(y)\cos(\beta) \tag{3}$$

The trailed vortices from this distribution of local lift will change the inflow, so that the angle should more accurately also be corrected with the induction as e.g. described for a wind turbine blade in [25]. A numerical test was made to estimate the influence of using the lift distribution calculated directly from Eq. (3) on the inflow and use this angle when projecting the pressure force to find a more accurate lift distribution. It was concluded that the effect on the final bound circulation distribution is quite small due to taking the cosine to the angle and therefore only the geometric angle is used when estimating the lift distribution from the pressure difference across the vortex generator.

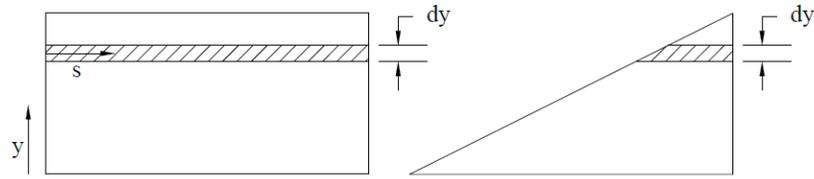

Figure 7 Integrating the pressure jump at the height y from the leading edge s=0 to the trailing edge s=c(y) for determining the load $f_n$(y) [N/m] normal to the VGs surface.

The bound circulation can now be computed from Eq. (1) as

$$\Gamma_b(y) = \frac{l(y)}{\rho u_x(y)} \qquad (4)$$

To fulfill Helmholtz's theorem, vortex filaments with strength equal to the gradient of the bound circulation must be trailed from the trailing edge as

$$d\Gamma_t = \gamma(y)dy \qquad (5)$$

where $\gamma$ is the distribution of bound circulation. However, after some short downstream distance these trailed vortices will curl up to one strong tip vortex with the strength

$$\Gamma_t = \int_{h_s}^{h} \gamma(y)d\,y = \int_{h_s}^{h} d\Gamma_b(y) \qquad (6)$$

so that the strength of the trailed vortex simply becomes the integral of the bound circulation from some starting value, $h_s$, to the tip y=h. If it had been a pure airplane wing, one should start the integration at y=0, but to avoid interference with the horizontal wall at y=0 the starting value is set to a fraction of the VG height and the value chosen was 0.2h. Looking at the streamlines in Figure 13, as well as from quantitative estimation, one can observe that the circulation for the innermost part of the VG does not contribute significantly to the tip vortex. To estimate the strength of the tip vortex from the computed flow in the wake of the VGs one can apply the best fit to a Lamb-Oseen Vortex (LOV) (in polar coordinates)

$$u_\theta(r) = \frac{\Gamma_{LOV}}{2\pi r}\left(1 - \exp\left(-\frac{r^2}{\varepsilon^2}\right)\right) \qquad (7)$$

$\varepsilon$ is a parameter that characterizes the radius of the vortex. This approach was first implemented for vortex generators in [26] and later applied in the works of others, see e.g. [27]. An alternative approach is to estimate the strength of the VG generated vortex by the path integral of the velocity component in the tangential direction (in polar coordinates) as

$$\Gamma_t = \int_0^{2\pi} u_\theta r d\theta \qquad (8)$$

Two constraints are taken into account for choosing the most representative radius r. First, the radius should be large enough to enclose all vorticity from one vortex and small enough not be influenced by the neighboring vortex. The hypothesis that the vortex strength can be estimated from the distribution of bound circulation on the VG as in Eq. (6) can now be tested by comparing it to the circulation behind the VG computed directly from the velocity in the vortex using Eqs. (7) and (8). If this hypothesis is valid, then one can estimate the expected strength from the boundary profile as the one shown in Figure 6 and assuming some airfoil data Cl(α) for the profiles used on the VG (often simply a thin flat plate) and where the angle of attack is approximately the pitch angle β.

## 4. Results and Discussion

Figures 8 and 9 display the obtained circulation in the wake found from Eqs. (7) and (8) for the rectangular and triangular VGs, respectively. Each subfigure of these two Figures compares the experimental results to the CFD results from one of the applied turbulence models, on the (left) Reynolds stress model, (middle) the k-ω SST and (right) the Spalart-Allmaras model. The path integrated circulation obtained from the experiments (o) and from the simulations (x) is also compared to the LOV circulation estimate (+). The circulation is calculated in five and two downstream planes for the rectangular (Figure 8) and triangular (Figure 9) configurations, respectively.

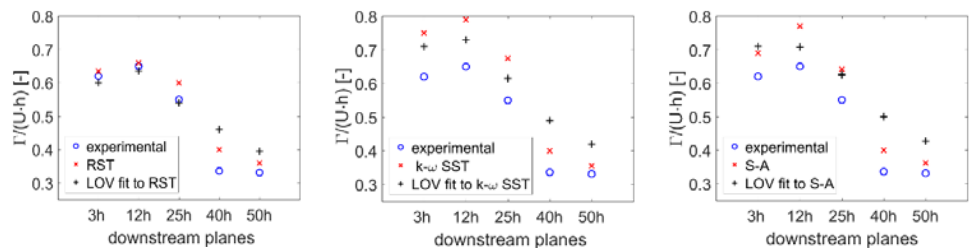

Figure 8 Circulation results for the rectangular configuration testing Reynolds Stress, k-ω SST, Spalart-Allmaras turbulence models in five downstream planes.

The experimentally determined circulation is computed from the (ensemble) averaged measurements/simulations, so meandering effects will impair on the resulting circulation as can be seen in the downstream development. The Reynolds - Stress Turbulence (RST) model (Figure 8, left) provides the most satisfactory comparison between the experimental [20,21] and the numerical results using the STAR-CCM+ model herein.

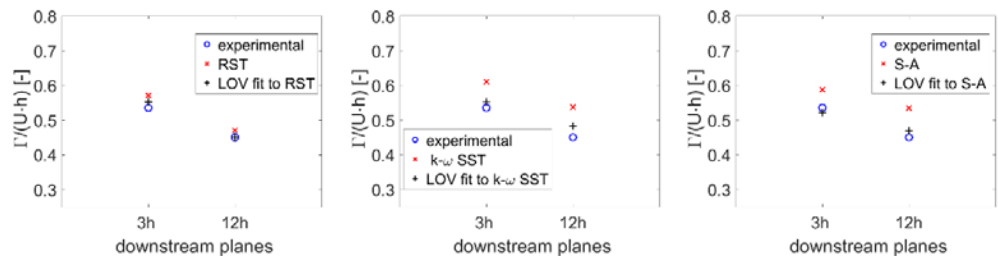

Figure 9 Circulation results for the triangular configuration testing Reynolds Stress, k-ω SST, Spalart-Allmaras turbulence models in two downstream planes.

For the triangular VG geometry, the circulation is investigated in two downstream planes as shown in Figure 9, since the ensemble averaged vortex cores in this setting are not always well distinguishable beyond 12h behind the VGs due to meandering [20,21]. For the first downstream plane, the triangular geometry yields a similar circulation compared to the rectangular while the RST turbulence model result is again the most accurate fit between the experimental and the numerical results. A better agreement is expected for the RST model since it includes correlations (flow history effects) and it is not based on an eddy viscosity model (as e.g. the two-equation and other lower order models). Furthermore, RST models can better capture the large-scale vortex motion through the inclusion of the material derivative of the $\overline{u'_i u'_j}$ correlations and it does not impose local isotropy or eddy viscosity in the same manner as lower order models [28,29]. The lower order models are widely known to not be able to capture large scale gradients, including stagnation, separation and rotation, which is in fact the hallmark of the flows produced by VGs.

As expected, the circulation estimated by the LOV is lower compared to the one estimated by the line integration due to the inherent axi-symmetry of the LOV approach (see Figure 10).

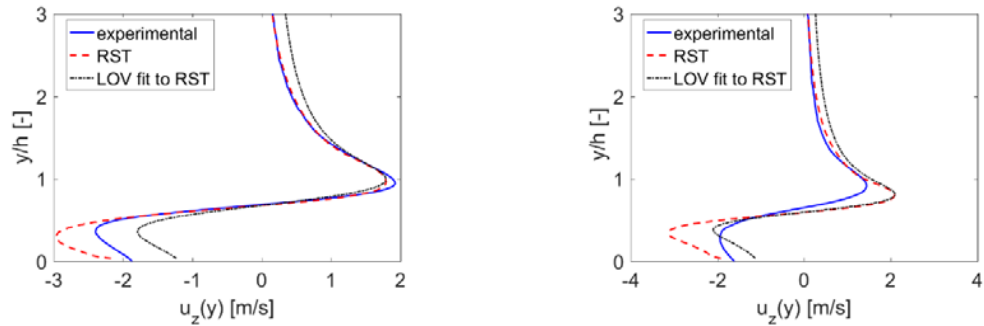

Figure 10 Determination of Γ with the use of Lamb - Oseen Vortex, (left) rectangular VG and (right) triangular VG, RST turbulence model.

Since the RST turbulence model provides the most accurate comparison between the experimental and numerical results for both geometries, it is used to validate the hypothesis for the estimation of the generated vortex strength from the spanwise distribution of bound circulation. This distribution along the VG height, obtained from the RST simulation, is depicted for both investigated geometries in Figure 11. As expected, the bound circulation weakens from the root to the tip of the VG height.

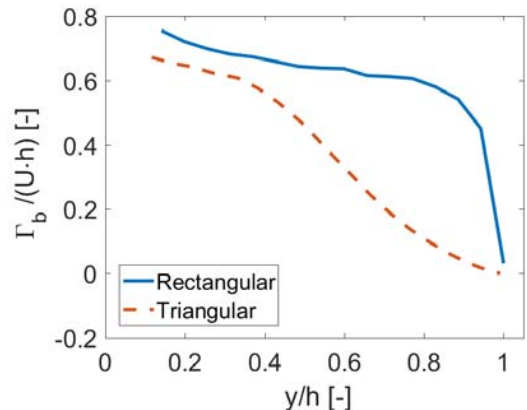

Figure 11 Distribution of the bound circulation along VGs height.

However, significant differences in bound circulation distribution of the two tested geometries can be observed. For the rectangular case (Figure 11, solid curve), the bound circulation displays an approximately flat distribution within a height range of y/h = [0.2:0.8], followed by a sharp drop due to the abrupt termination in VG geometry near the top. In the triangular VG case (Figure 11, dashed curve), the slope decreases approximately linearly across the main part of the VG, which is expected due to the linearly varying chord length with VG height. More careful inspection reveals an inflection point (the curvature changes sign) at approximately y/h≈0.6. This behavior is clearer in the circulation of the trailed filaments, shown in Figure 12 , which are estimated from the gradient of the bound circulation. Here, the effect is instead manifested in a shift in the slope of the trailed circulation at y/h≈0.6.

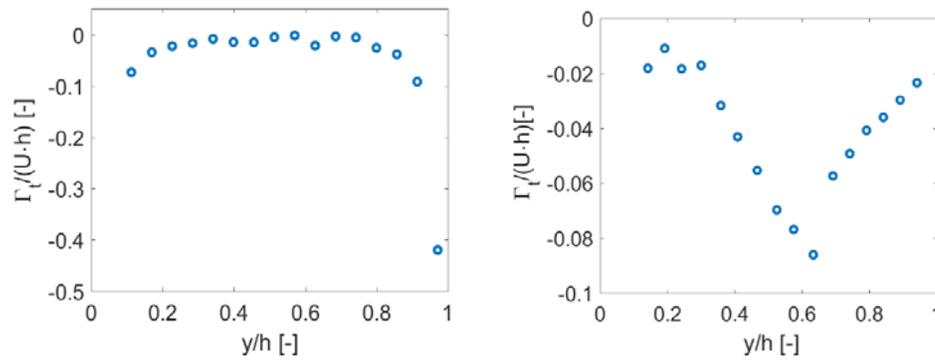

Figure 12 Distribution of the trailed circulation along VGs height. Rectangular VG (left) and triangular VG (right).

For the rectangular VGs (Figure 12, left), the trailed circulation is strongest near the tip since the spanwise gradient is the highest. This indicates that the downstream vortex is mostly generated near the tip of the vane. For the triangular VGs (Figure 12, right), the highest magnitude of the trailed circulation is observed in the vicinity of the inflection point, while most of the vortex filaments contribute to the circulation of the downstream vortex in contrast to the rectangular ones. This can be explained by the fact that vorticity is accumulated along the entire inclined edge of the triangular vane in order to form the vortex into the wake, as can be seen in Figure 13 .

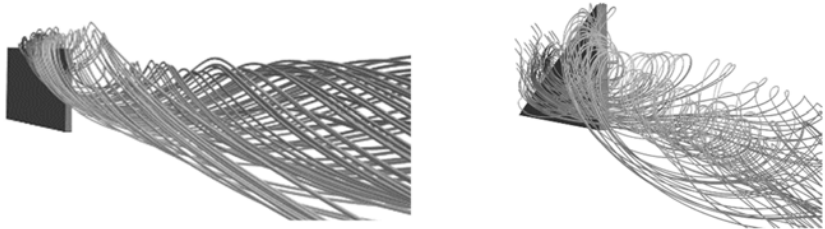

Figure 13 Trailing vortex associated streamlines visualized for the (left) rectangular and (right) triangular vortex generators, RST turbulence model.

The formation of the trailing vortex is thus not only occurring near the tip, but in fact with the collection of the released trailing vortex filaments occurring along the entire edge and rolling up to one strong vortex as seen in Figure 13 (right). This may also explain why the vortex cores formed behind triangular VGs are sometimes significantly more distorted from axi-symmetry than those generated from rectangular VGs (c.f. [9,30]). For the rectangular VGs, the summation of the circulation of the filaments between y=0.2h and the tip results in a numerical value $\Gamma_t/U\cdot h$=0.65, which is almost identical with the value obtained from the line integration and the LOV, see Figure 8. The trailed circulation for the triangular VG from integrating the bound circulation is equal to $\Gamma_t/U\cdot h$=0.61 and still quite close to value of 0.55 estimated from the wake flow using LOV and line integration, which further supports the hypothesis.

## 5. Conclusions

In the current investigations, PIV measurements and RANS simulations are used to verify a hypothesis that the strength of the streamwise vortex behind a VG responsible for mixing the boundary layer can be found from the distribution of bound circulation that again in an engineering model may be estimated from 2-D airfoil data assuming local 2-D flow at different heights from the root of the VG. Such an approach has been suggested by Poole et al. [31], where they formally applied a classical lifting line model to model the strength of the generated vortex from several VG geometries and compared this to measurements by Wendt [32]. Depending on how the bound circulation was estimated from airfoil theory the errors varied between 5 to 25%. In this paper the bound circulation at various wall-normal distances along the VG was determined by CFD and not by airfoil data and from this distribution the trailed vortices were determined in the same way as in a classical lifting line model and a very good agreement with the values computed from the wake velocities was found. Also, the trailed circulation from the CFD fits well with the experimentally obtained trailed circulation and the trends are consistent for both the classical rectangular and triangular VG shapes. All this supports the idea of making an engineering approach that can estimate the strength of the generated streamwise vortex based on the upstream boundary layer profile and the VG geometry, where the spanwise bound circulation then is determined from some assumed 2-D airfoil data, for the various profiles used along the VGs. Knowing the circulation distribution on the VG one can therefore expect to estimate quite accurately the strength of the trailed vortex. It is, however, not the purpose of this paper to refine such a model, but to verify that the assumptions behind this approach are valid. It could even be possible to implement higher order models of this kind, e.g. considering time variations in the flow to improve the prediction accuracy. Also the spanwise distribution of the bound circulation is, as expected, quite different behind a triangular shaped VG than that behind a rectangular VG in the same boundary layer resulting in a different vortex structure as seen in Figure 12 and 13.

### References

1. Taylor H.D., The Elimination of Diffuser Separation by Vortex Generators, Research Department Report No. R-4012-3, United Aircraft Corporation, East Hartford, Connecticut, **1947**.
2. Taylor H.D., Design Criteria for and Applications of the Vortex Generator Mixing Principle, Research Department Report No. M-15038-1, United Aircraft Corporation, East Hartford, Connecticut, **1948.**
3. Taylor H.D., Summary Report on Vortex Generators, Research Department Report No. R-05280-9, United Aircraft Corporation, East Hartford, Connecticut, **1950**
4. Haipeng Wang, Bo Zhang, Qinggang Qiu, Xiang Xu. Flow control on the NREL S809 wind turbine airfoil using vortex generators. *Energy* **2017**, Volume 118, Pages 1210-1221.
5. H.H. Xia, G.H. Tang, Y. Shi, W.Q. Tao. Simulation of heat transfer enhancement by longitudinal vortex generators in dimple heat exchangers. *Energy* **2014**, Volume 74, pages 27-36.
6. Lei Luo, Wei Du, Songtao Wang, Lei Wang, Xinhong Zhang. Multi-objective optimization of a solar receiver considering both the dimple/protrusion depth and delta-winglet vortex generators. *Energy* **2017**, Volume 137, pages 1-19.
7. Ting Ma, Jaideep Pandit, Srinath V. Ekkad, Scott T. Huxtable, Qiuwang Wang. Simulation of thermoelectric-hydraulic performance of a thermoelectric power generator with longitudinal vortex generators. *Energy* **2015**, Volume 84, pages 695-703.
8. Amin Ebrahimi, Farhad Rikhtegar, Amin Sabaghan, Ehsan Roohi. Heat transfer and entropy generation in a microchannel with longitudinal vortex generators using nanofluids. *Energy* **2016**, Volume 101, pages 190-201.
9. Velte, C.M., Hansen M.O.L, and Okulov, V. Helical structure of longitudinal vortices embedded in turbulent wall-bounded flow. *Journal of Fluid Mechanics* **2009**, 619, pp. 167-177.
10. Pearcey, H.H. Shock induced separation and its prevention by design and boundary layer control, in: G.V. Lachmann (Ed.), Boundary Layer and Flow Control, vol. 2, Pergamon Press, **1961**, pp. 1166–1344.
11. Jones, J.P. The Calculation of Paths of Vortices from a System of Vortex Generators, and a Comparison with Experiments. Aeronautical Research Council, Technical report C.P. No. 361, **1957**.
12. Smith, F.T. Theoretical prediction and design for vortex generators in turbulent boundary layers. *J. Fluid Mech.* **1994**, 270, pp. 91–131.

13. Lamb H. Hydrodynamics, 6th ed. Cambridge Univ. Press, Cambridge, **1932**.

14. Hansen M.O.L., Westergaard C. Phenomenological Model of vortex generators, in: IEA, Aerodynamics of Wind Turbines, 9th Symposium, **1995**, pp. 1–7.

15. Chaviaropoulos, P.K., Politis, E.S., Nikolaou, I.G. Phenomenological modeling of vortex generators. in: Proc. EWEC 2003, 16–19 June, Madrid, Spain, **2003**.

16. Velte, C.M., Hansen, M.O.L, Okulov, V. Multiple vortex structures in the wake of a rectangular winglet in ground effect. *Experimental Thermal and Fluid Science* **2016**, vol: 72, pp. 31–39

17. Charalampous, A., Foucaut, J.M., Cuvier, C., Velte, C.M., Hansen, M.O.L., Verification of numerical modeling of VG flow by comparing to PIV data, Wind Energy Science Conference 2017, Kgs. Lyngby, DTU, Denmark, June 26-29, 2017.

18. Bender E.E, Anderson B.H. and Yagle P.J. Vortex generator modeling for Navier Stokes codes. Proceedings of the 1999 3rd ASME, FEDSM'99-6929, San Francisco, California, USA, 18-23 July **1999**

19. Manolesos M., Sørensen N.N., Troldborg N., Florentie, L., Papadakis, G., Voutsinas, S. Computing the flow past Vortex Generators: Comparison between RANS Simulations and Experiments. The Science of Making Torque from Wind (TORQUE **2016**), IOP Publishing Journal of Physics: Conference Series 753.

20. Velte, C.M., Braud, C., Coudert S. and Foucaut J.-M. Vortex generator induced flow in a high re boundary layer. *Journal of Physics: Conference Series (online)* **2014**, vol. 555, no. 1, p. 012102.

21. Foucaut, J M et al. Influence of light sheet separation on SPIV measurement in a large field spanwise plane. *Measurement Science and Technology* **2014**. 25(3).

22. STAR CCM+ Users Manual. http://www.cd-adapco.com/products/star-ccm/documentation

23. Fernández-Gámiz, U.,Velte, C.M, Réthoré, P.E, Sørensen, N. N.; Egusquiza, E. Testing of self-similarity and helical symmetry in vortex generator flow simulations *In: Wind Energy, Vol.* 19, No. 6, 2016, p. 1043–1052.

24. Velte, C.M., Hansen, M.O.L., and Cavar D. Flow analysis of vortex generators on wing sections by stereoscopic particle image velocimetry measurements. *Published 24 January 2008, IOP Publishing Ltd. Environmental Research Letters, Volume 3, Number 1.*

25. Rahimi, H., Schepers J.G., Shen W.Z., Ramos Garcia, N., Scneider M.S., Micallef D., Simao Ferreira C.J., Jost E., Klein L., Herráez I., Evaluation of different methods for determining the angle of attack on wind turbine blades with CFD results under axial inflow conditions, *Renewable Energy*, **2018**, (125) pp.866-876.

26. Velte, C.M., Hansen, M.O.L.,Okulov, V.L., Helical structure of longitudinal vortices embedded in turbulent wall-bounded flow. Volume 619 25 January **2009**, pp. 167-177. DOI: https://doi.org/10.1017/S0022112008004588

27. Martínez-Filgueira, P., Fernandez-Gamiz, U., Zulueta,E., Errasti, I., Fernandez-Gauna, B., Parametric study of low-profile vortex generators. *International Journal of Hydrogen Energy*, ISSN: 0360-3199, Vol: 42, Issue: 28, Page: 17700-17712, **2017**. http://doi.org/10.1016/j.ijhydene.2017.03.102.

28. Launder, B., Reece, G., & Rodi, W. Progress in the development of a Reynolds-stress turbulence closure. *Journal of Fluid Mechanics* **1975**, *68*(3), pp. 537-566.

29. Gibson, M., & Rodi, W. A Reynolds-stress closure model of turbulence applied to the calculation of a highly curved mixing layer. *Journal of Fluid Mechanics* **1981**, *103*, pp.161-182.

30. Velte, C. M., Hansen, M. O. L., & Cavar, D. Flow analysis of vortex generators on wing sections by stereoscopic particle image velocimetry measurements. *Environmental Research Letters* **2008**, *3*(1).

31. Poole, D.J., Bevan R.L.T., Allen, C.B., Rendall, T.C.S. An Aerodynamic Model for Vane-Type Vortex Generators, 8th AIAA Flow Control Conference, **2016**

32. Wendt, J. Parametric study of vortices shed from airfoil vortex generators. AIAA Journal, 42(11):2185-2195, 2004.